\documentclass[journal]{IEEEtran}

\usepackage{enumerate}
\usepackage{graphicx}
\usepackage{epsf}
\usepackage{subfigure}
\usepackage{amsmath}
\usepackage{amssymb}
\usepackage{array}
\usepackage{setspace}
\usepackage[amsmath,thmmarks]{ntheorem}
\usepackage[ruled,lined]{algorithm2e}
\usepackage{algorithmic}
\usepackage{color}
\usepackage{amsfonts}
\usepackage{dsfont}
\usepackage{xfrac}
\usepackage{breqn}
\usepackage{bbm}
\usepackage{cite}

\graphicspath{{Fig/},{fig/}}

\theoremseparator{.}

\DeclareMathOperator{\diag}{diag}

\begin{document}

\title{\huge RIS-Aided Fluid Antenna Array-Mounted UAV Networks}


\author{
Li-Hsiang Shen,~\IEEEmembership{Member,~IEEE},
Yi-Hsuan Chiu}

\maketitle

\begin{abstract}
This paper investigates reconfigurable intelligent surface (RIS)-assisted unmanned aerial vehicle (UAV) downlink networks with fluid antennas (FA), where RIS enables non-line-of-sight (NLoS) transmissions. Moreover, the FA is equipped on the UAV offering dynamic antenna position adjustment, enhancing spatial diversity besides UAV deployment. We aim at total downlink rate maximization while ensuring minimum user rate requirement. We consider joint optimization of active UAV beamforming, passive RIS beamforming, UAV deployment and FA position adjustment. To address the complex problem, we propose beamfomring for RIS/UAV and FA-UAV deployment (BRAUD) scheme by employing alternative optimization, successive convex approximation (SCA) and sequential rank-one constraint relaxation (SROCR) method for the decomposed subproblems. Simulation results demonstrate the effectiveness of RIS-FA-UAV, achieving the highest rate among existing architectures without FA/UAV/RIS deployment and without proper beamforming. Moreover, BRAUD achieves the highest rate among benchmarks of drop-rank method, heuristic optimizations and conventional zero-forcing beamforming as well as random method.

\end{abstract}

{\let\thefootnote\relax\footnotetext
{Li-Hsiang Shen and Yi-Hsuan Chiu are with the Department of Communication Engineering, National Central University, Taoyuan 320317, Taiwan. (email: shen@ncu.edu.tw and claire90428@gmail.com)}}

\section{Introduction}

Unmanned aerial vehicles (UAVs) have captured significant attention as a promising technology to create resilient and adaptable wireless communication networks. UAV-assisted communication networks not only extend coverage to remote or underserved areas but also boost energy efficiency (EE) through optimized deployment strategies \cite{1}. However, challenges emerge in complex or heavily obstructed environments, like urban areas with tall buildings or dense forests, where signal propagation losses can greatly impair communication quality, resulting in prevalent non-line-of-sight (NLoS) paths. Although higher deployment altitudes can improve line-of-sight (LoS) connectivity, it will introduce higher path loss. Striking a balance between altitude and signal integrity is essential yet challenging, as both LoS access and minimal path loss should be considered.

To address these challenges, reconfigurable intelligent surfaces (RIS) have emerged as an innovative solution \cite{6,7,acm}. By dynamically adjusting elemental phase-shifts and amplitudes, RIS beamforming can control reflection and scattering of wireless signals. As a crucial benefit, RIS greatly mitigates multipath fading, enhances system spectral efficiency, and expands potential service coverage for multiusers \cite{dstar,8}. Furthermore, the integration of UAVs with RIS marks a significant leap forward, enhancing the performance of emerging networks \cite{2,3}. Studies in \cite{4,5} have shown that the UAV-RIS combination significantly boosts system performance by maximizing sum rates at users while minimizing total power consumption. These advancements are achieved through optimized UAV deployment/beamforming, scheduling, RIS configurations, and power allocation.

\begin{figure}[!t]
\centering
\includegraphics[width=2.8in]{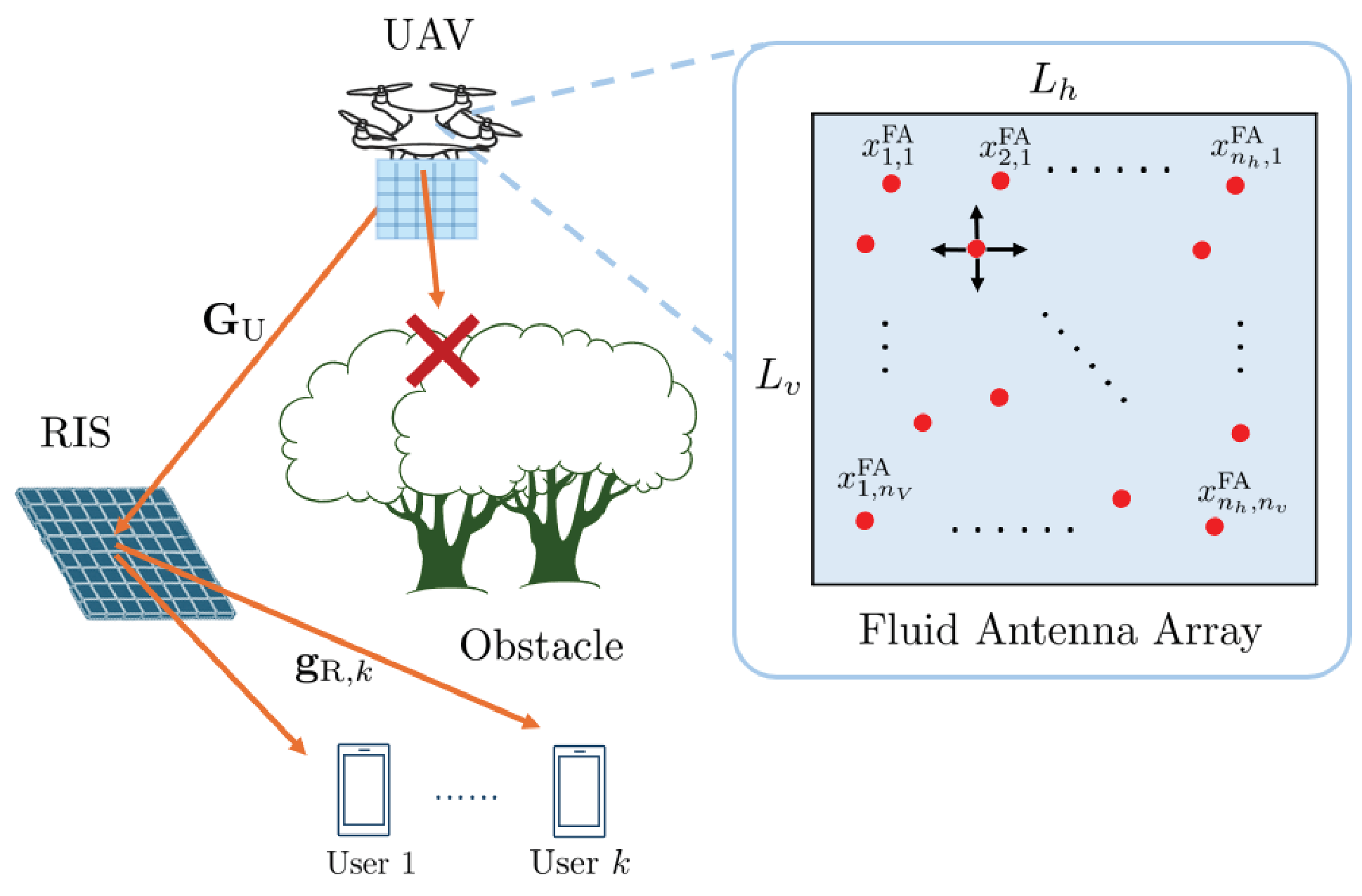}
\caption{The proposed architecture of RIS-aided FA-UAV downlink networks.} \label{architecture}
\end{figure}

Complementing UAV-RIS, fluid antenna (FA) systems \cite{9}, known as \textit{movable} antennas, are gaining substantial attention thanks to adapting wireless channels. Unlike UAV deployment operating on larger space, FA exploits fine-grained and real-time adjustment by physically repositioning the antenna element, introducing new degrees of freedom \cite{10}. This flexibility allows FA to dynamically respond to channel conditions and optimize signal paths. Hardware architecture and channel characterization for FA systems are studied in \cite{MAA}. FA systems highlight several advantages, including enhanced simultaneous multi-beamforming and low-power consumption. Lower transmission power is consumed than traditional fixed-position antenna arrays, as beamformed signals are better aligned to small-scale fading channel \cite{11}. Accordingly, the integration of FA systems holds significant potential for improving adaptability and efficiency in high-density or highly-interfered environments \cite{MAA}. In this work, we explore the potential of equipping UAVs with FA. UAV deployment compensates for large-scale fading, while FA position adjustments help mitigate small-scale fading more effectively than UAV deployment alone. RIS is incorporated to address potential transmission path blockages and further enhance multiuser service coverage. The main contributions of this paper are summarized as follows:
\begin{itemize}
    \item We propose a novel architecture deploying RIS in FA-mounted UAV networks, where RIS enables NLoS transmission. UAV deployment combats large-scale channel fading, while FA alleviates small-scale fading, introducing additional degrees of freedom through hybrid-scaled spatial diversity.
    
    \item We aim at maximizing the downlink sum rate while ensuring minimum user rate. We jointly optimize UAV beamforming, RIS beamforming, UAV deployment, and FA position. The unsolvable problem is decomposed into four subproblems solved by alternating optimization. We propose beamfomring for RIS/UAV and FA-UAV deployment (BRAUD) scheme by utilizing successive convex approximation (SCA) and sequential rank-one constraint relaxation (SROCR) method to convert them into solvable forms.

    \item Numerical results demonstrate that BRAUD achieves the highest rates compared to cases without FA, UAV, RIS or proper beamforming. Increasing the number of FA antennas and RIS elements enhances flexibility and degrees of freedom, further combating channel fading. The proposed BRAUD scheme also achieves the highest rate compared to benchmarks of drop-rank, heuristic optimization, conventional beamforming, and random method.
\end{itemize}

\section{System Model and Problem Formulation}
In Fig. \ref{architecture}, we consider the UAV serving $K$ downlink users equipped with a single antenna, with user set $\mathcal{K} = \{1, ...,k,..., K\}$. The UAV is equipped with an FA array having $N = N_h \cdot N_v$ transmit antenna elements. Notations $N_h$ and $N_v$ indicate the numbers of horizontal and vertical elements, respectively, with sets as $\mathcal{N}_h = \{1, ..., n_h, ..., N_h\}$ and $\mathcal{N}_v = \{1, ..., n_v, ..., N_v\}$. The size of FA is given by $L = L_h \cdot L_v$, where $L_h$ and $L_v$ are the lengths of horizontal and vertical sides. The RIS is composed of $M$ elements with phase-shift matrix $\mathbf{\Theta} = \diag \left(  e^{j \theta_1},..., e^{j \theta_m}, ...,  e^{j \theta_M} \right)$. Notation of $\theta_m$ indicates the phase-shift of the $m$-th RIS element. The 3D Cartesian coordinates are considered, i.e., positions of the RIS, UAV, and user $k$ are defined as $\mathbf{x}^{\text{R}} = [x^{\text{R}}, y^{\text{R}}, z^{\text{R}}]^T$, $\mathbf{u} = [x^{\text{U}}, y^{\text{U}}, z^{\text{U}}]^T$, and $\mathbf{x}_k^{\text{UE}} = [x_k^{\text{UE}}, y_k^{\text{UE}}, z_k^{\text{UE}}]^T$, respectively. Due to UAV mobility, the boundary of flying area is constrained by
\begin{align}
	\mathbf{u}_{\text{min}} \preceq \mathbf{u} \preceq \mathbf{u}_{\text{max}}, \label{uav_con2}
\end{align}
where $\mathbf{u}_{\text{min}}$ and $\mathbf{u}_{\text{max}}$ are the corresponding minimum and maximum coordinates of the flight boundary. Moreover, due to velocity limitation, UAV movement must comply with
\begin{align}
    \| \mathbf{u} - \mathbf{u}' \| \leq \delta, \label{uavmove}
\end{align}
where $\mathbf{u}'$ is the previous location, whilst $\delta$ is the moving distance threshold. The $(n_h, n_v)$-th element of the FA array response vector $\mathbf{A}^{\text{FA}}\in \mathbb{C}^{n_h \times n_v}$ can be expressed as
\begin{align} \label{FAR}
[\mathbf{A}^{\text{FA}}]_{n_h,n_v} = e^{j \mathbf{k} \mathbf{x}_{n_h, n_v}^{\text{FA}}}, 
\end{align}
where the wave vector is $\mathbf{k} = \frac{2\pi}{\lambda} \left[ 
\cos \left( \psi \right) \cos(\vartheta), 
\cos \left( \psi\right) \sin(\vartheta), 
\sin \left( \psi \right)
\right]$, $\lambda$ is wavelength, and the position of each FA element is $\mathbf{x}_{n_h, n_v}^{\text{FA}} = [x^{\text{FA}}_{n_h, n_v}, y^{\text{FA}}_{n_h, n_v},z^{\text{FA}}_{n_h, n_v}]^T$. Notations of $\psi$ and $\vartheta$ are azimuth and elevation angles of UAV. Due to hardware limitation, FA element is geometrically confined by the following constraints:
\begingroup
\allowdisplaybreaks
\begin{subequations} \label{FA_con}
\begin{align}
	& x^{\text{FA}}_{n_h,n_v} - x^{\text{FA}}_{n_{h-1},n_v'} \geq d_{x,\text{th}},  \quad \forall n_h\in \mathcal{N}_h, n_v, n_v' \in \mathcal{N}_v, \label{FA_con1}
	\\
	& y^{\text{FA}}_{n_h,n_v} - y^{\text{FA}}_{n_h',n_{v-1}} \geq d_{y,\text{th}}, \quad \forall n_h, n_h'\in \mathcal{N}_h, n_v\in \mathcal{N}_v, \label{FA_con2}
	\\
	&  x^{\text{FA}}_{1,n_v} \geq 0, \ 
	   x^{\text{FA}}_{N_h,n_v} \leq L_h, \quad \forall n_v\in \mathcal{N}_v, \label{FA_con3}
	\\
	&  y^{\text{FA}}_{n_h,1} \geq 0, \ 
	   y^{\text{FA}}_{n_h,N_v} \leq L_v, \quad \forall n_h\in \mathcal{N}_h, \label{FA_con4}
	\\
	& z^{\text{FA}}_{n_h,n_v} = z_{n'_h,n'_v}, \notag \\ & \qquad \forall n_h, n'_h\in \mathcal{N}_h, \ n_v, n'_v \in \mathcal{N}_v, \ n_h \neq n'_h, n_v\neq n'_v. \label{FA_con5}
\end{align}
\end{subequations}
\endgroup
Constraints \eqref{FA_con1} and \eqref{FA_con2} avoid collision by guaranteeing inter-element distance thresholds $d_{x,\text{th}}$ and $d_{y,\text{th}}$ in horizontal and vertical directions, respectively. Constraint \eqref{FA_con3} confines the range of FA, i.e., the position of the first element should be larger than $0$, whereas the last one cannot exceed the horizontal width $L_h$. Similarly, constraint for vertical direction can be deduced in \eqref{FA_con4}. Constraint \eqref{FA_con5} exhibits the identical height of FA.


The channel follows Rician fading between UAV and RIS as
\begin{align} \label{CH_UAV}
\mathbf{G}_{\text{U}} = \sqrt{h_0 d_{\text{U}}^{-\kappa}} \left( P_{\text{U}}^{\text{LoS}} \cdot \mathbf{A} + P_{\text{U}}^{\text{NLoS}} \cdot \mathbf{G}_{\text{U}}^{\text{NLoS}} \right), 
\end{align}
where $\mathbf{A}= {\rm vec}(\mathbf{A}_{\text{R}})^H {\rm vec}(\mathbf{A}^{\text{FA}}) \in \mathbb{C}^{M \times N} $ is the array response matrix of UAV to RIS, where $\mathbf{A}_{\text{R}}$ for RIS can be identically obtained based on \eqref{FAR}. ${\rm vec}(\cdot)$ reshapes a matrix into a vector representation. Notation of $h_0 = \frac{c}{4\pi d_0 f}$ indicates the pathloss at the reference distance $d_0$, $c$ is the light speed, and $f$ is the operating frequency. Notation $d_{\text{U}}$ indicates the distance between UAV and RIS, and $\kappa$ is the pathloss exponent. The LoS probability between UAV and RIS is given by $P_{\text{U}}^{\text{LoS}} = \frac{1}{1 + a_1 \cdot e^{a_2 (\vartheta - a_1)}}$ \cite{uav}, where $a_1$ and $a_2$ are constants determined by the environments. Note that $P_{\text{U}}^{\text{NLoS}} = 1 - P_{\text{U}}^{\text{LoS}}$ is the probability of NLoS. Moreover, NLoS path $\mathbf{G}_{\text{U}}^{\text{NLoS}} \in \mathbb{C}^{M \times N}$ follows Rayleigh fading. Similarly, the channel between RIS and user $k$ is
\begin{align}
\mathbf{g}_{\text{R},k} = \sqrt{h_0 d_{\text{R},k}^{-\kappa}} \left( P_{\text{R},k}^{\text{LoS}} \cdot \mathbf{g}_{\text{R},k}^{\text{LoS}} + P_{\text{R},k}^{\text{NLoS}} \cdot \mathbf{g}_{\text{R},k}^{\text{NLoS}} \right), 
\end{align}
where the pertinent parameters can follow similar definitions to those in \eqref{CH_UAV}. Accordingly, the effective RIS channel can be obtained as $\mathbf{g}_{k} = \mathbf{g}_{\text{R},k}^H \mathbf{\Theta} \mathbf{G}_{\text{U}} \in \mathbb{C}^{1 \times N }$. The transmitted signal and the beamforming vector of the $k$-th user are defined as $s_{k}$ and $ \mathbf{w}_{k} \in \mathbb{C}^{N \times 1}$, respectively. Then, the received signal of user $k$ is given by
\begin{align} \label{r_signall}
y_{k} &= \mathbf{g}_{k} \mathbf{w}_{k} s_{k} 
+ \mathbf{g}_{k} \sum_{k' \in \mathcal{K} \backslash k} \mathbf{w}_{k'} s_{k'} + n_{k},
\end{align}
where $ n_{k} \sim \mathcal{CN}(0, \sigma_{k}^2) $ denotes complex additive white Gaussian noise (AWGN) with power $\sigma_{k}^2$. Therefore, the signal-to-interference-plus-noise ratio (SINR) is obtained as
\begin{align} \label{sinr}
\gamma_{k} = 
\frac{\left| \mathbf{g}_{k} \mathbf{w}_{k} \right|^2}{
    \sum_{k' \in \mathcal{K} \backslash k} \left| \mathbf{g}_{k} \mathbf{w}_{k'} \right|^2  
    + \sigma_{k}^2}.
\end{align}
Based on \eqref{sinr}, the achievable rate of user $k$ can be obtained as $R_{k} = \log_2 (1 + \gamma_{k})$.

We intend to investigate the rate maximization problem with the arguments of transmit beamforming $\mathbf{w}_{k}$, RIS phase-shift configuration $\boldsymbol{\Theta}$, deployment of UAV $\mathbf{u}$ and position of FA elements $\mathbf{X}^{\rm FA}$, which can be represented by
\begingroup
\allowdisplaybreaks
 \begin{subequations} \label{problem_T}
\begin{align}
    &\mathop{\max}\limits_{\substack{  \mathbf{w}_{k},\boldsymbol{\Theta}, \mathbf{u}, \mathbf{X}^{\rm FA}}} \quad  
     \sum_{k\in \mathcal{K}} R_{k} \label{obj} \\
     & \text{s.t. } \quad \theta_{m} \in [0, 2\pi), \quad \forall m\in \mathcal{M}, \label{con1} \\
     &\quad\quad \sum_{k\in \mathcal{K}} \|\mathbf{w}_k\|^2 \leq P_{\text{max}}, \label{con2} \\ 
     &\quad\quad R_{k} \geq R_{\text{min}}, \quad \forall k\in \mathcal{K}, \label{con3} \\
     &\quad\quad \mathbf{u} \in \mathcal{D}_{X}, \label{con5} \\
     &\quad\quad \mathbf{X}^{\rm FA} \in \mathcal{R}_{X}. \label{con4}
\end{align}
 \end{subequations}
  \endgroup
Constraint \eqref{con1} limits RIS phase-shifts, whereas \eqref{con2} guarantees maximum transmit power $P_{\text{max}}$. Constraint \eqref{con3} ensures the minimum required user rate, whilst \eqref{con5} restricts UAV deployment $\mathcal{D}_{X}=\{\eqref{uav_con2}, \eqref{uavmove} \}$. Constraint \eqref{con4} confines FA deployment of $\mathcal{R}_{X} = \{\eqref{FA_con1}-\eqref{FA_con5}\}$. The primary challenge of this problem lies on nonconvexity, nonlinearity, and coupled variables. To address these issues, we employ the alternating optimization to iteratively optimize the decomposed in the subsequent section.

\section{Proposed BRAUD Algorithm}

\subsection{Transmit Beamforming Optimization}
 
   Firstly, we prioritize optimizing the subproblem of transmit beamforming $\mathbf{W} = [\mathbf{w}_1, \mathbf{w}_2,...,\mathbf{w}_K]$ with the
fixed parameters $\{ \boldsymbol{\Theta}, \mathbf{u},\mathbf{X}^{\text{FA}}\}$. Then, the subproblem can be formulated as
\begin{align} \label{problem_Beamforming}
    &\mathop{\max}\limits_{\substack{  \mathbf{W}}} \
     \sum_{k\in \mathcal{K}} R_{k}(\mathbf{W}) \quad \text{s.t. } \ \eqref{con2}, \eqref{con3}.
\end{align}  
We transform problem \eqref{problem_Beamforming} by introducing the auxiliary variables $\mathbf{F}_k \triangleq \mathbf{w}_k \mathbf{w}_k^H$, with the rank-one constraint as ${\rm{rank}}(\mathbf{F}_k) = 1$. The total solution set of transmit beamforming as $\mathbf{F} \triangleq \left[ \mathbf{F}_1, \mathbf{F}_2, \dots, \mathbf{F}_K \right]$. Then, the objective function can be reformulated as $R_{k}(\mathbf{F}) = \log_2 \left( 1 + \frac{{\rm{Tr}}(\mathbf{F}_k \mathbf{g}_k^H \mathbf{g}_k)}{\sum_{j\in \mathcal{K} \backslash k } {\rm{Tr}}\left( \mathbf{F}_{j} \mathbf{g}_k^H \mathbf{g}_k \right) + \sigma_k^2} \right)$. Then, constraint $\eqref{con3}$ can be rewritten as
\begin{align} \label{r_signal}
	{\rm{Tr}} \left(\mathbf{F}_k \mathbf{g}_k^H \mathbf{g}_k  \right)  + \left( 1 - 2^{R_{\text{min}}} \right) \left( \sum_{j\in \mathcal{K} \backslash k } {\rm{Tr}}\left( \mathbf{F}_j \mathbf{g}_k^H \mathbf{g}_k \right) + \sigma_k^2 \right) \geq 0.
\end{align}
Additionally, it can be observed that $\mathbf{F}_k$ is semi-definite. Accordingly, the subproblem \eqref{problem_Beamforming} can be transformed into
\begingroup
\allowdisplaybreaks
 \begin{subequations} \label{problem_Beamforming01}
\begin{align}
    &\mathop{\max}\limits_{\substack{  \mathbf{F}}} \quad  
     \sum_{k\in \mathcal{K}} R_{k}(\mathbf{F}) \label{obj_B} \\
     & \text{s.t. }  
     \quad \eqref{r_signal},
     \quad \mathbf{F}_k \succeq 0 , \  \forall k\in \mathcal{K}, 
     \quad \sum_{k\in\mathcal{K}} {\rm{Tr}}(\mathbf{F}_k) \leq P_{\text{max}}, \label{conB1} \\
     & \qquad\ {\rm{rank}}(\mathbf{F}_k) = 1, \quad  \forall k\in \mathcal{K}. \label{conB3}
\end{align}
 \end{subequations}
  \endgroup
The transformed problem \eqref{problem_Beamforming01} is still non-convex due to the objective function \eqref{obj_B}, constraint \eqref{r_signal} and rank-one constraint \eqref{conB3}. Firstly, we employ the method of difference of two concave functions (D.C.) to tackle the rate term. We rewrite $R_{k}(\mathbf{F})=f_k(\mathbf{F}) - z_k(\mathbf{F})$, where $f_k(\mathbf{F}) =  \log_2 \left( \sum_{j\in \mathcal{K}} {\rm Tr} \left( \mathbf{F}_j \mathbf{g}_k^H \mathbf{g}_k \right) + \sigma_k^2 \right)$, and $z_k(\mathbf{F}) = \log_2 \left( \sum_{j\in \mathcal{K} \backslash k } {\rm Tr} \left( \mathbf{F}_j \mathbf{g}_k^H \mathbf{g}_k \right) + \sigma_k^2 \right)$. However, due to non-convex term of $z_k(\mathbf{F})$, we adopt SCA with the first-order Taylor approximation as $\hat{z}_k(\mathbf{F}) \approx z_k(\mathbf{F}^{(t)}) + {\rm Tr}\left( \nabla z_k^T (\mathbf{F}^{(t)}) (\mathbf{F} -\mathbf{F}^{(t)}) \right)$, where
$
\nabla z_k^T (\mathbf{F}^{(t)}) = 
\frac{ \mathbf{g}_k^H \mathbf{g}_k }{ \sum_{j\in \mathcal{K} \backslash k } \left({\rm Tr} \left( \mathbf{F}_j^{(t)} \mathbf{g}_k^H \mathbf{g}_k \right) + \sigma_k^2  \right) \cdot \ln 2}.
$
Till now, the objective \eqref{obj_B} and constraint \eqref{r_signal} become convex. Secondly, we employ SROCR to solve rank-one constraint \eqref{conB3}. Unlike the conventional semi-definite relaxation (SDR) method that directly drops it, SROCR can gradually relax \eqref{conB3} \cite{srocr}. At the $t$-th iteration, rank-one constraint \eqref{conB3} can be updated as
\begin{align}
    \left( \boldsymbol{\lambda}_k^{(t)} \right)^H \mathbf{F}_k \boldsymbol{\lambda}_k^{(t)} \geq \tau_{1,k}^{(t)} {\rm{Tr}}\left( \mathbf{F}_k \right), \label{18}
\end{align}
where $\tau_{1,k}^{(t)} \in [0, 1]$ is a relaxation constant, indicating the ratio of the largest eigenvalue of $\mathbf{F}_k^{(t)}$ to ${\rm{Tr}}( \mathbf{F}_k^{(t)} )$ at previous iteration $t$. Notation of $\boldsymbol{\lambda}_k^{(t)}$ represents the eigenvector associated with the eigenvalue of $\mathbf{F}_k^{(t)}$. As a result, we can rewrite problem \eqref{problem_Beamforming01} as 
\begingroup
\allowdisplaybreaks
\begin{align} \label{problem_Beamforming02}
    &\mathop{\max}\limits_{\substack{  \mathbf{F}}} \quad  
     \sum_{k\in \mathcal{K}} \left( f_k(\mathbf{F}) - \hat{z}_k(\mathbf{F}) \right) \quad \text{s.t. } \eqref{conB1}, \eqref{18}. 
\end{align}
\endgroup
The problem \eqref{problem_Beamforming02} is convex and a semidefinite programming (SDP) problem, which can be solved by convex optimization tools. After $\mathbf{F}_k$ is obtained, Cholesky decomposition \cite{16} is used to reconstruct the $\mathbf{w}_k$, which is omitted for brevity.

\subsection{RIS Configuration Optimization}
After solving $\mathbf{W}$, we proceed to solve the subproblem with respect to (w.r.t.) RIS configuration $\boldsymbol{\Theta}$ as 
\begingroup
\allowdisplaybreaks
\begin{align} \label{problem_theta1}
     &\mathop{\max}\limits_{\substack{  \mathbf{\Theta}}} \
     \sum_{k\in \mathcal{K}} R_{k}(\mathbf{\Theta})   \quad
      \text{s.t. } \ \eqref{con1}, \eqref{con3}. 
\end{align}
\endgroup
The term in $R_k({\mathbf{\Theta}})$ can be rewritten as $\left| \mathbf{g}_k \mathbf{w}_k \right|^2 
	= \left| \mathbf{g}_{\text{R},k}^H \mathbf{\Theta} \mathbf{G}_{\text{U}} \mathbf{w}_k \right|^2 
	\overset{\underset{(a)}{}}{=} \left| \mathbf{\boldsymbol{\varphi}}^H \left( \diag(\mathbf{g}_{\text{R},k}) \mathbf{G}_{\text{U}} \mathbf{w}_k \right) \right|^2  \overset{\underset{(b)}{}}{=} \left| \boldsymbol{\varphi}^H {\mathbf{q}}_{k,(k)} \right|^2 
	\overset{\underset{(c)}{}}{=} \boldsymbol{\varphi}^H \mathbf{Q}_{k,(k)} \boldsymbol{\varphi}$, where (a) follows $\boldsymbol{\varphi} = \left[ e^{j \theta_1} , ... , e^{j \theta_M} \right]^T$, whereas (b) defines ${\mathbf{q}}_{k,(k)} = \diag(\mathbf{g}_{\text{R},k}) \mathbf{G}_{\text{U}} \mathbf{w}_k$. Last equality (c) defines $\mathbf{Q}_{k,(k)} = {\mathbf{q}}_{k,(k)} {\mathbf{q}}_{k,(k)}^H$. Then, we introduce the auxiliary variable $\mathbf{V} = \boldsymbol{\varphi} \boldsymbol{\varphi}^H$. The sum rate can be rewritten as
\begin{align}
        R_k(\mathbf{V}) = \log_2\left(1 + \frac{{\rm{Tr}}(\mathbf{V} \mathbf{Q}_{k,(k)})}{\sum_{j\in \mathcal{K} \backslash k} {\rm{Tr}}(\mathbf{V} \mathbf{Q}_{k,(j)}) + \sigma_k^2}\right). \label{ratev}
\end{align}
From \eqref{ratev}, the constraint \eqref{con3} can be derived as
\begin{align}
    {\rm{Tr}}(\mathbf{V} \mathbf{Q}_{k,(k)}) + \left(1 - 2^{R_{\text{min}}}\right) \left( \sum_{j\in \mathcal{K} \backslash k} {\rm{Tr}}(\mathbf{V} \mathbf{Q}_{k,(j)}) + \sigma_k^2 \right) \geq 0. \label{qosv}
\end{align}
The subproblem \eqref{problem_theta1} is transformed to
\begingroup
\allowdisplaybreaks
 \begin{subequations} \label{problem_theta2}
\begin{align}
    &\mathop{\max}\limits_{\substack{  \mathbf{V}}} \quad  
     \sum_{k\in \mathcal{K}} R_{k}(\mathbf{V})   \\
     & \text{s.t. } 
     \quad \eqref{qosv},
     \ \mathbf{V} \succeq 0, 
     \ {\rm{Tr}}(\mathbf{V}) = M, 
     \ {\rm{rank}}(\mathbf{V}) = 1,  \label{vcon1}
\end{align}
 \end{subequations}
  \endgroup
where $\mathbf{V}$ satisfies the semi-definite properties in constraints \eqref{vcon1}. The non-convex problem \eqref{problem_theta2} can be solved by SROCR following the previous subproblem, with rank-one constraint replaced by
\begin{align}
    \left( \boldsymbol{\xi}^{(t')} \right)^H \mathbf{V} \boldsymbol{\xi}^{(t')} \geq \tau_2^{(t')} {\rm{Tr}}(\mathbf{V}), \label{vrank1}
\end{align}
where $\tau_{2}^{(t')} \in [0, 1]$ is the ratio of the largest eigenvalue of $\mathbf{V}_k^{(t')}$ to ${\rm{Tr}}( \mathbf{V}_k^{(t')} )$ at previous iteration $t'$. Notation of $\boldsymbol{\xi}_k^{(t')}$ represents the eigenvector associated with the eigenvalue of $\mathbf{V}_k^{(t')}$. The objective function in problem \eqref{problem_theta2} can be transformed into the difference of two concave functions, i.e., $R_{k}(\mathbf{V}) = f_k(\mathbf{V})- z_k(\mathbf{V})$, where $f_k(\mathbf{V}) = \log_2 \left( \sum_{j \in \mathcal{K}} {\rm Tr}(\mathbf{V} \mathbf{Q}_{k,(j)}) + \sigma_k^2 \right)$ and $z_k(\mathbf{V}) = \log_2 \left( \sum_{j \in \mathcal{K} \backslash k} {\rm Tr}(\mathbf{V} \mathbf{Q}_{k,(j)}) + \sigma_k^2 \right)$. Employing a first-order Taylor approximation for non-convex term of $z_k(\mathbf{V})$ yields the problem as
\begingroup
\allowdisplaybreaks
 \begin{subequations} \label{problem_theta3}
\begin{align}
    &\mathop{\max}\limits_{\substack{  \mathbf{V}}} \quad  
     \sum_{k\in \mathcal{K}} \left( f_k(\mathbf{V}) - \hat{z}_k(\mathbf{V}) \right)  \\
     & \text{s.t. } \quad \eqref{qosv}, \eqref{vrank1}, \ \mathbf{V} \succeq 0, 
     \ {\rm{Tr}}(\mathbf{V}) = M, \nonumber 
\end{align}
 \end{subequations}
  \endgroup
where $ \hat{z}_k(\mathbf{V}) = z_k\left(\mathbf{V}^{(t')}\right) + \nabla z_k^T\left(\mathbf{V}^{(t')}\right) \left( \mathbf{V} - \mathbf{V}^{(t')} \right)  $ and $\nabla z_k^T \left(\mathbf{V}^{(t')} \right)
 = \frac{\sum_{j\in \mathcal{K} \backslash k} \mathbf{Q}_{k,(j)} }{\left( \sum_{j\in \mathcal{K} \backslash k} {\rm{Tr}}( \mathbf{V}^{(t')} \mathbf{Q}_{k,(j)}) + \sigma_k^2 \right) \cdot \ln 2}$. Problem \eqref{problem_theta3} has been transformed into an SDP problem, which can be effectively solved by arbitrary convex optimization tools.
  
\subsection{UAV Deployment Optimization}
    Given the solved parameters of $\{\mathbf{W},\mathbf{\Theta}\}$ and fixed $\mathbf{u}$, the problem of UAV deployment can be reformulated as
    \begingroup
\allowdisplaybreaks
\begin{align} \label{problem_UAV1}
    &\mathop{\max}\limits_{\substack{  \mathbf{u}}} \  
     \sum_{k\in \mathcal{K}} R_{k}(\mathbf{u})   \quad \text{s.t.} \ \eqref{con3}, \eqref{con5}. 
\end{align}
\endgroup
The sum rate can be written as
\begingroup
\allowdisplaybreaks
\begin{align}
&R_k(\mathbf{u}) = \log_2 \left( \sum_{j\in \mathcal{K}} \left| h_0 d_{\text{R},k}^{-\frac{\kappa}{2}} d_{\text{U}}^{-\frac{\kappa}{2}} \overline{\mathbf{g}}_{\text{R},k} \mathbf{\Theta} \overline{\mathbf{G}}_{\text{U}} \mathbf{w}_j \right|^2 + \sigma_k^2 \right) \nonumber \\
& \quad - \log_2 \left( \sum_{j\in \mathcal{K} \backslash k} \left| h_0 d_{\text{R},k}^{-\frac{\kappa}{2}} d_{\text{U}}^{-\frac{\kappa}{2}} \overline{\mathbf{g}}_{\text{R},k} \mathbf{\Theta}  \overline{\mathbf{G}}_{\text{U}} \mathbf{w}_j \right|^2 + \sigma_k^2 \right), \label{urate}
\end{align}
\endgroup
where $\overline{\mathbf{g}}_{\text{R},k} = P_{\text{R},k}^{\text{LoS}} \cdot \mathbf{g}_{\text{R},k}^{\text{LoS}} + P_{\text{R},k}^{\text{NLoS}} \cdot \mathbf{g}_{\text{R},k}^{\text{NLoS}}$ and $\overline{\mathbf{G}}_{\text{U}} = P_{\text{U}}^{\text{LoS}} \cdot \mathbf{A} + P_{\text{U}}^{\text{NLoS}} \cdot \mathbf{G}_{\text{U}}^{\text{NLoS}}$. The square term in $R_k(\mathbf{u})$ can be rewritten as 
$h_0^2 d_{\text{R},k}^{-\kappa} d_{\text{U}}^{-\kappa} \left( \overline{\mathbf{g}}_{\text{R},k} \mathbf{\Theta} \overline{\mathbf{G}}_{\text{U}} \mathbf{w}_j \right)^H \left( \overline{\mathbf{g}}_{\text{R},k} \mathbf{\Theta} \overline{\mathbf{G}}_{\text{U}} \mathbf{w}_j \right) \triangleq C_{k,j} D_k, 
$
where 
$C_{k,j} = h_0^2 \left( \overline{\mathbf{g}}_{\text{R},k} \mathbf{\Theta} \overline{\mathbf{G}}_{\text{U}} \mathbf{w}_j \right)^H \left( \overline{\mathbf{g}}_{\text{R},k} \mathbf{\Theta} \overline{\mathbf{G}}_{\text{U}} \mathbf{w}_j \right)$
and
$D_k = d_{\text{R},k}^{-\kappa} d_{\text{U}}^{-\kappa}$. Accordingly, constraint \eqref{con3} can be rewritten as
\begin{align}
    \sum_{j\in \mathcal{K}} C_{k,j} D_k + \left(1 - 2^{R_{\text{min}}}\right) \left( \sum_{j\in \mathcal{K} \backslash k} C_{k,j} D_k + \sigma_k^2 \right) \geq 0. \label{uavqos}
\end{align}
We further recast the objective function as
\begin{align} \label{problem_UAV2}
    \mathop{\max}\limits_{\substack{  \mathbf{u}}} \
     \sum_{k\in \mathcal{K}} \left( f_k(\mathbf{u}) - z_k(\mathbf{u}) \right)  \quad \text{s.t.} \ \eqref{con5},\eqref{uavqos},
\end{align}
where $f(\mathbf{u}) = \log_2 \left( \sum_{k\in \mathcal{K}} D_k C_{k,j} + \sigma_k^2 \right) $ and $z(\mathbf{u}) = \log_2 \left( \sum_{\substack{j\in \mathcal{K} \backslash k}} D_k C_{k,j} + \sigma_k^2 \right) $. To handle the coupled term, we rewrite $D_k$ as
\begin{align} \label{rrr}
 D_k = d_{\text{R},k}^{-\kappa} d_{\text{U}}^{-\kappa} \triangleq q_{k} \cdot p = (q_{k}+p)^2/2 - q_k^2/2 - p^2 /2,
\end{align}
where auxiliary variables are defined as $q_{k} \triangleq d_{\text{R},k}^{-\kappa}$ and $p \triangleq d_{\text{U}}^{-\kappa}$, where $\left( |\mathbf{x}^{\text{R}} - \mathbf{x}_k^{\text{UE}}|^2  \right)^{\frac{\kappa}{2}} = q_k^{-1}$ and $\left( |\mathbf{u} - \mathbf{x}^{\text{R}}|^2 \right)^{\frac{\kappa}{2}} = p^{-1}$.
Consequently, we further approximate the non-convex parts by employing SCA. Thus, problem \eqref{problem_UAV2} is reformulated as
 \begingroup
\allowdisplaybreaks
 \begin{subequations} \label{problem_UAV3}
\begin{align}
    &\mathop{\max}\limits_{\substack{  \mathbf{u}, D_k, q_k, p}} \quad  
     \sum_{k\in \mathcal{K}} \left( f_k \left(\mathbf{u} \right) - \hat{z}_k (\mathbf{u})   \right) \\
     \text{s.t. } & \quad \eqref{con5},\eqref{uavqos}, \nonumber \\
     & \quad \left( |\mathbf{x}^{\text{R}} - \mathbf{x}_k^{\text{UE}}|^2  \right)^{\frac{\kappa}{2}} \leq 1/q_k^{(l)} - 1/q_k^{(l)^2} (q_k - q_k^{(l)}),  \ \forall k \in \mathcal{K}, \label{r1}\\
     & \quad \left( |\mathbf{u} - \mathbf{x}^{\text{R}}|^2 \right)^{\frac{\kappa}{2}} \leq 1/p^{(l)} - 1/p^{(l)^2} (p - p^{(l)}), \label{r2}\\
     & \quad D_{k} \geq (q_k + p)^{2}/2 - \left[ q_k^{(l)^2}/2 + q_k^{(l)} \left( q_k - q_k^{(l)} \right) \right] \nonumber \\     
     &\hspace{16mm} - \left[ p^{(l)^2}/2 + p^{(l)} \left( p - p^{(l)} \right) \right], \ \forall k \in \mathcal{K}, \label{r3}
\end{align}
 \end{subequations}
  \endgroup
where $\hat{z}_k(\mathbf{u}) = z_k(\mathbf{u}^{(l)}) + \nabla z_k^T(\mathbf{u}^{(l)}) (\mathbf{u} - \mathbf{u}^{(l)})$ and $\nabla z_k^T (\mathbf{u}) = \frac{\sum_{j\in\mathcal{K} \backslash k} C_{k,j}}{ \left( \sum_{j\in\mathcal{K} \backslash k} D_k C_{k,j} + \sigma_k^2 \right) \cdot \ln 2 }$. At right hand side of constraints \eqref{r1}--\eqref{r3}, SCA is employed for $q_k^{-1}$, $p^{-1}$ and \eqref{rrr} w.r.t. $q_{k}$, $p$ and joint $\{q_{k}, p\}$ respectively. Note that superscript $(l)$ indicates the solution of UAV deployment obtained at the previous $l$-th iteration. That is, the next UAV position is optimized and predicted based on the current channel and position information. Then, problem \eqref{problem_UAV3} becomes convex and solvable.

\subsection{FA Position Matrix Optimization}
Given the solved parameters of $\{\mathbf{W}, \boldsymbol{\Theta}, \mathbf{u}\}$, the optimization of FA adjustment can be reformulated as
\begin{align} \label{problem_FA1}
  	\mathop{\max}\limits_{\substack{  \mathbf{x}^{\rm FA} }} \
     \sum_{k\in \mathcal{K}} R_{k}(\mathbf{x}^{\rm FA})  \quad  \text{s.t.} \ \eqref{con3}, \eqref{con4}.
\end{align}
Note that $R_{k}(\mathbf{x}^{\rm FA})$ follows the same form as \eqref{urate}. We rewrite the square term in \eqref{sinr} as $\left| \mathbf{g}_{\text{R},k}^H \mathbf{\Theta} \mathbf{G}_{\text{U}} \mathbf{w}_j \right|^2 
	= \left| \mathbf{g}_{\text{R},k}^H \mathbf{\Theta} \sqrt{h_0 d_{\text{U}}^{-\kappa}} \! \left( P_{\text{U}}^{\text{LoS}}  \mathbf{A} \!+\! P_{\text{U}}^{\text{NLoS}} \mathbf{G}_{\text{U}}^{\text{NLoS}} \right) \! \mathbf{w}_j \right|^2 
	=  h_0 d_{\text{U}}^{-\kappa} \left| \mathbf{m}_{1,k} \mathbf{A} \mathbf{w}_j  + m_{2,k,j}  \right|^2$,
where $\mathbf{m}_{1,k} = P_{\text{U}}^{\text{LoS}} \cdot \mathbf{g}_{\text{R},k}^H \mathbf{\Theta} $ and $m_{2,k,j} =  P_{\text{U}}^{\text{NLoS}} \cdot \mathbf{g}_{\text{R},k}^H \mathbf{\Theta} \mathbf{G}_{\text{U}}^{\text{NLoS}} \mathbf{w}_j$. Subsequently, problem \eqref{problem_FA1} can be reformulated as
\begin{align}\label{problem_FA2_0}
    &\mathop{\max}\limits_{\substack{  \mathbf{x}^{\rm FA}}}  
     \sum_{k\in \mathcal{K}} \left( f_k(\mathbf{x}^{\rm FA}) - z_k(\mathbf{x}^{\rm FA}) \right)  \quad \text{s.t.} \ \eqref{con3}, \eqref{con4},  
\end{align}
where $f_k(\mathbf{x}^{\rm FA}) \!=\! \log_2 \left( \sum_{j\in \mathcal{K}} h_0 d_{\text{U}}^{-\kappa} \left| \mathbf{m}_{1,k} \mathbf{A} \mathbf{w}_j \!+\! m_{2,k,j} \right|^2 \!+\! \sigma_k^2 \right)$
and
$z_k(\mathbf{x}^{\rm FA}) \!=\! \log_2 \left( \sum_{j\in \mathcal{K} \backslash k} h_0 d_{\text{U}}^{-\kappa} \left|  \mathbf{m}_{1,k} \mathbf{A} \mathbf{w}_j \!+\! m_{2,k,j} \right|^2 \!+\! \sigma_k^2 \right)
$. We observe that $f_k(\mathbf{x}^{\rm FA})$, $z_k(\mathbf{x}^{\rm FA})$ and \eqref{con3} are non-convex due to complex variables in the exponential term. Then, we acquire the following transformed problem by SCA as
\begingroup
\allowdisplaybreaks
 \begin{subequations} \label{problem_FA2}
\begin{align}
    & \mathop{\max}\limits_{\substack{  \mathbf{x}^{\rm FA}}} \quad  
    \sum_{k\in \mathcal{K}} \left( \hat{f}_k(\mathbf{x}^{\rm FA}) -  \hat{z}_k(\mathbf{x}^{\rm FA}) \right)   \\
     &\text{s.t. } 
     \quad \eqref{con4},\quad \hat{f}_k(\mathbf{x}^{\rm FA}) -  \hat{z}_k(\mathbf{x}^{\rm FA}) \geq R_{\text{min}}, \forall k \in \mathcal{K},
\end{align}
 \end{subequations}
  \endgroup
  where
$\hat{f}_k(\mathbf{x}^{\rm FA}) = f_k(\mathbf{x}^{{\rm FA} (l')}) + {\rm Tr}\left(\nabla f_k^T(\mathbf{x}^{{\rm FA} (l')}) (\mathbf{x}^{\rm FA} - \mathbf{x}^{{\rm FA} (l')}) \right)$
and
$\hat{z}_k(\mathbf{x}^{\rm FA}) = z_k(\mathbf{x}^{{\rm FA} (l')}) + {\rm Tr}\left( \nabla z_k^T(\mathbf{x}^{{\rm FA} (l')}) (\mathbf{x}^{\rm FA} - \mathbf{x}^{{\rm FA} (l')}) \right)$. Note that the detailed calculation of first-order gradients of $\nabla f_k(\mathbf{x}^{{\rm FA}})$ and $\nabla z_k(\mathbf{x}^{{\rm FA}})$ are omitted for brevity. To this end, problem \eqref{problem_FA2} becomes convex and can be readily solved by arbitrary convex optimization tools. Overall algorithm of the proposed BRAUD scheme is conducted by iteratively solving $\mathbf{W}$ in problem \eqref{problem_Beamforming02}, solving $\boldsymbol{\Theta}$ in problem \eqref{problem_theta3}, solving $\mathbf{u}$ in problem \eqref{problem_UAV3}, and solving $\mathbf{X}^{\text{FA}}$ in problem \eqref{problem_FA2}. The total computational complexity order of BRAUD is given by $\mathcal{O}\left(  \left[ N^{4.5} + M^4 \sqrt{N} + M^{3.5} + (MK)^3 \right] \cdot T \log_2(\frac{1}{\varepsilon}) \right)$, where $T$ indicates the upper bound of the total iteration and $\varepsilon$ indicates the accuracy for the approximation of the interior-point method. Note that four terms in the complexity order respectively represent the four subproblems in \eqref{problem_Beamforming02}, \eqref{problem_theta3}, \eqref{problem_UAV3}, and \eqref{problem_FA2}.


\section{Simulation result}

%
%
%


Simulation results are provided to validate the effectiveness of RIS-FA-UAV and proposed BRAUD scheme. The altitudes of UAV/RIS/users are set to $\{100,30,1.7\}$ m. Other related parameters are set as follows: $a_1 = 0.3$, $a_2 = 0.7$, $\kappa = 2.2$, $\delta = 20$ m, $\lambda = 0.1$ m, $P_{\text{max}} = 2$ Watt, $R_{\text{min}} = 1$ bps/Hz. Users are uniformly and randomly distributed within a radius of $\{100, 300\}$ meters for scenarios 1 and 2, respectively. The RIS is deployed at a distance of $\left[100, 500\right]$ meters away from the users.

\begin{figure}
	\centering
\begin{minipage}[t]{0.24\textwidth}
	\centering
	\includegraphics[width=1.8 in]{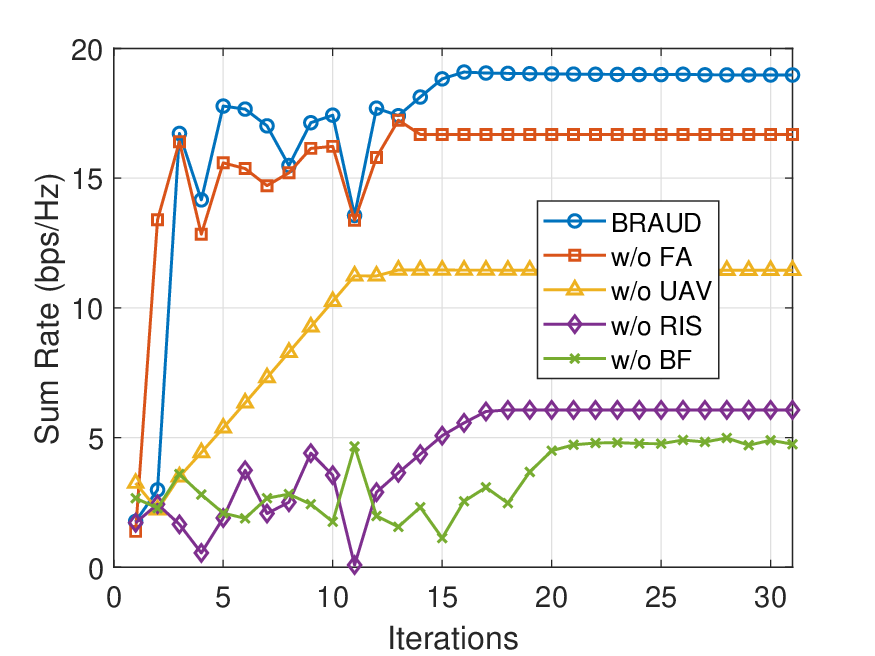}
\caption{Convergence of BRAUD.} \label{convergence}
\end{minipage}
\begin{minipage}[t]{0.24\textwidth}
	\centering
	\includegraphics[width=1.8 in]{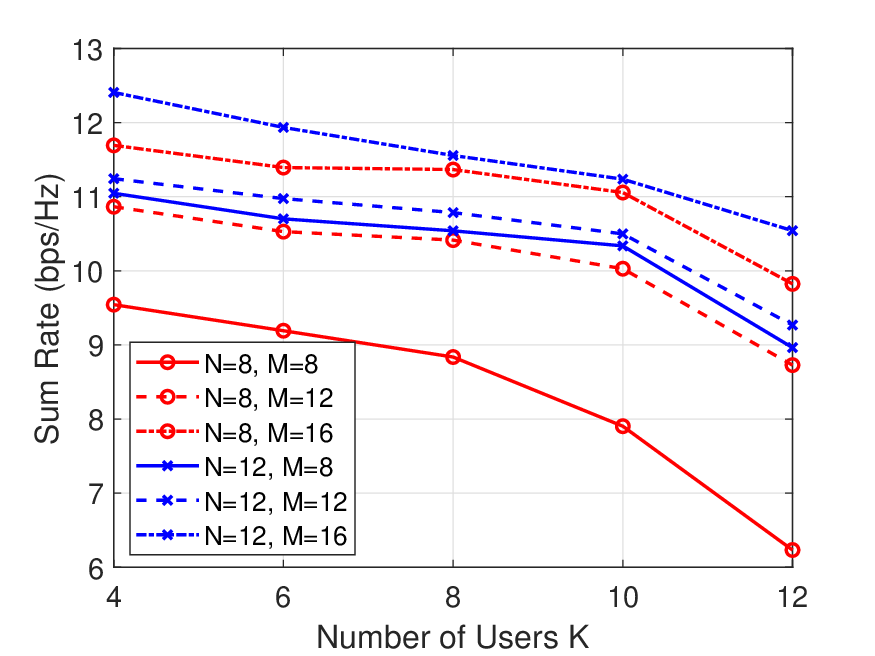}
\caption{Rate of different $\{K,M,N\}$.} \label{antenna}
\end{minipage}	
\\	
\begin{minipage}[t]{0.24\textwidth}
	\centering
	\includegraphics[width=1.8 in]{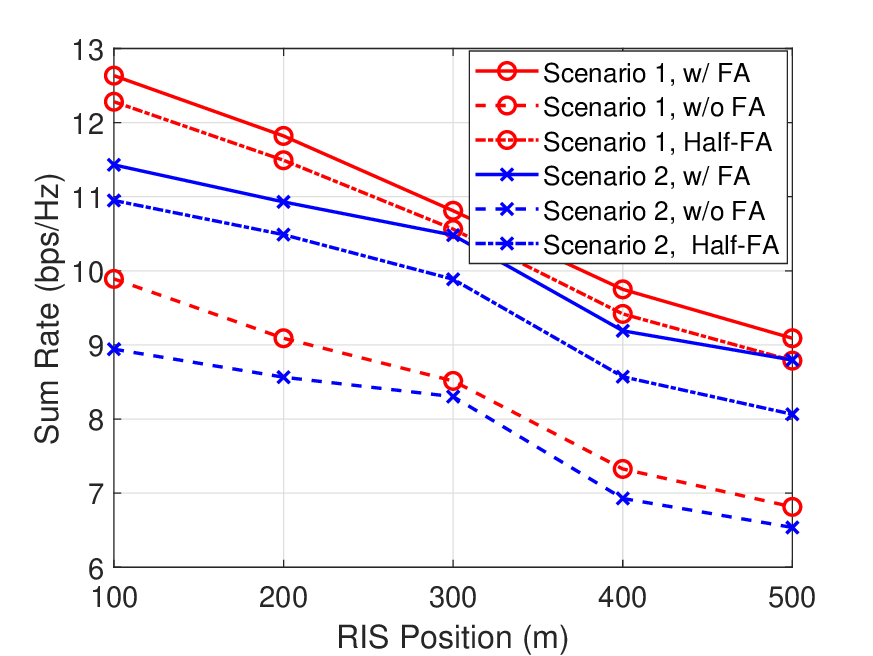}
\caption{Different RIS/FA deployment.} \label{risfa}
\end{minipage}	
\begin{minipage}[t]{0.24\textwidth}
	\centering
	\includegraphics[width=1.8 in]{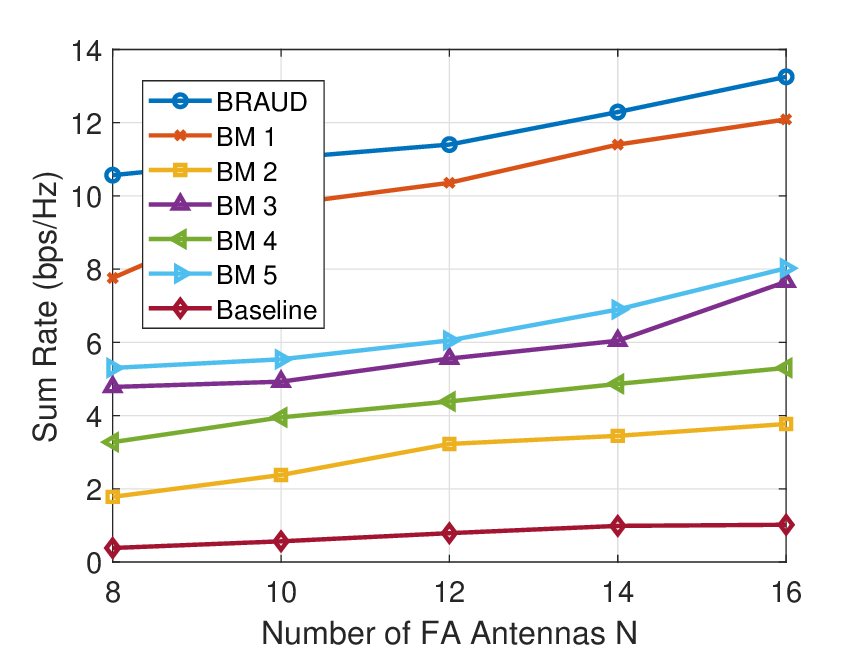}
\caption{Benchmark comparison.} \label{benchmark}
\end{minipage}		
\end{figure}

In Fig. \ref{convergence}, we compare BRAUD across different cases without FA, UAV, RIS optimization, and random beamforming (w/o BF). Notably, in the case of fixed UAV deployment (w/o UAV), the sum rate converges smoothly due to the absence of distance-based channel fading variations. The worst rate performance is observed with improper UAV/RIS beamforming, but the rate improves when only RIS beamforming is suboptimal. Active/passive beamforming can partially compensate for the fixed UAV deployment. By comparing the cases w/o FA and w/o UAV, UAV deployment becomes more critical, as it mitigates large-scale fading. Overall, BRAUD achieves substantial rate improvements of around $13\%$, $65\%$, $213\%$, and $287\%$ compared to cases without FA, UAV, RIS deployment, and beamforming, respectively.

Fig. \ref{antenna} illustrates the rate performance with different numbers of users $K$, FA antennas $N$ and RIS elements $M$. We observe that the rate decreases with more users increases, due to insufficient spatial diversity. However, more FA antennas provide additional adjustable positions to combat small-scale fading in the UAV-RIS link. Similarly, increasing RIS elements enhances rate by enabling more phase-shift adjustments for different users, mitigating channel fading in the RIS-user link. Also, with more FA antennas, the requirement for RIS elements can be reduced. For example, less rate improvement is observed when increasing $M$ from $8$ to $16$ when $N=12$, compared to that when $N=8$, because FA can adjust both antenna position and active beamforming.

Fig. \ref{risfa} compares rates for different RIS deployments and FA adjustments under scenarios 1 (hotspot) and 2 (sparse users). Here, w/ FA (full-FA) indicates that all antennas are movable, while w/o FA represents fixed position with evenly spaced antennas. Half-FA assumes that $50\%$ antennas are movable. We observe that deploying the RIS closer to users yields a higher rate due to stronger reflected signals. Scenario 1 shows a higher rate primarily due to lower propagation loss compared to scenario 2. Additionally, the half-FA configuration results in only a $4\%$ rate reduction compared to full-FA, achieving a compelling balance between potential hardware complexity/cost and system performance.

In Fig. \ref{benchmark}, we compare BRAUD with four methods: Benchmark (BM) 1 applies a drop-rank approach \cite{drop}, ignoring rank-one constraints in beamforming. BM 2 uses zero-forcing beamforming without accounting for interference. BM 3 employs a heuristic genetic algorithm (GA) \cite{GA}, incorporating bio-inspired operations of elite selection, crossover and mutation. BMs 4 and 5 adopt multi-armed bandit (MAB) with quantized parameters \cite{mab} and continuous MAB, respectively. Note that the baseline randomizes all parameters. Zero-forcing effectively adjusts both active and passive beamforming, yielding improved rate over the baseline. However, GA and MAB, with their heuristic mechanisms, can better optimize deployment and configuration, resulting in higher rate than zero-forcing. Moreover, continuous MAB without quantization errors can have higher rate than quantized MAB and GA. Due to potential local solutions, drop-rank achieves a higher rate than BMs 3-5. Additionally, BRAUD outperforms BM 1 by a significant margin, as the drop-rank approach leads to relaxed constraints that result in less precise parameter recovery.

\section{Conclusions}
We have investigated the potential of RIS-FA-UAV network for maximizing the total downlink rate while maintaining user rate requirement. The proposed BRAUD scheme iteratively solves the subproblems w.r.t. active UAV beamforming, passive RIS beamforming, UAV deployment and FA position adjustment. The employment of SCA and SROCR addresses the challenges of nonconvexity, nonlinearity, and rank-one constraints. Numerical results have revealed BRAUD achieves the significant rate improvement of around $13\%$, $65\%$, $213\%$, and $287\%$ compared to architectures without FA, UAV, RIS deployment, and without proper beamforming, respectively. Benefited from additional FA/RIS elements, BRAUD achieves higher rates due to the increased degrees of freedom for mitigating channel fading. Furthermore, BRAUD outperforms existing benchmarks, including the drop-rank method, zero-forcing beamforming, heuristic optimization, and baseline random methods, in terms of rate.

\linespread{0.8}
\bibliographystyle{IEEEtran}
\bibliography{IEEEabrv}
\end{document}